\documentclass[12pt]{article}
\usepackage{epsf}

\begin{document}

\newcommand{\gtrsim}{ \mathop{}_{\textstyle \sim}^{\textstyle >} }
\newcommand{\lesssim}{ \mathop{}_{\textstyle \sim}^{\textstyle <} }

\newcommand{\rem}[1]{{\bf #1}}

\renewcommand{\thefootnote}{\fnsymbol{footnote}}
\setcounter{footnote}{0}
\begin{titlepage}

\def\thefootnote{\fnsymbol{footnote}}

\begin{center}

\hfill TU-745\\
\hfill hep-ph/0505252\\
\hfill May, 2005\\

\vskip .75in

{\Large \bf 
Reconstructing Dark Matter Density with 
$e^+e^-$ Linear Collider
in Focus-Point Supersymmetry
}

\vskip .75in

{\large
Takeo Moroi, Yasuhiro Shimizu and Akira Yotsuyanagi
}

\vskip 0.25in

{\em
Department of Physics, Tohoku University,
Sendai 980-8578, JAPAN}

\end{center}

\vskip .5in

\begin{abstract}

It has been known that, in the focus point scenario of supersymmetry,
the thermal relic of the lightest superparticle (LSP) is known to be a
good candidate of the cold dark matter.  Assuming that the LSP in the
focus-point scenario be the cold dark matter, we address a question
how and how well the relic density of the LSP can be determined once
the superparticles are found at future $e^+e^-$ linear collider. We
will see that the determinations of the mass of the LSP as well as
those of the Higgsino-like chargino and neutralinos, which will be
possible by a study of the decay kinematics of the chargino or by
threshold scan, will give us important information to theoretically
reconstruct the relic density.  Even if the Higgsino-like
superparticles and the LSP are the only superparticles which are
kinematically accessible, relic density of the LSP may be calculated
with the accuracy of factor $\sim 2$; by adopting a mild theoretical
assumption or by determining the masses of the Wino-like
superparticles, uncertaintiy can be reduced to $\sim 10\ \%$ or
smaller.

\end{abstract}

\end{titlepage}

\renewcommand{\thepage}{\arabic{page}}
\setcounter{page}{1}
\renewcommand{\thefootnote}{\#\arabic{footnote}}
\setcounter{footnote}{0}

High energy physics and cosmology have had very important connections
for deeper understandings in each field.  In particular, progresses in
physics at the energy frontier sometimes greatly improved our
knowledges on the history of the universe.  One of the most famous
examples is the big-bang nucleosynthesis \cite{BBN_Gamow}; with the
understandings of the interactions among the light elements and the
expansion of the universe by general relativity, it became clear that
the light nuclei were synthesized in the early universe.  Remarkably,
nowadays, light-element abundances can be precisely calculated and,
for ${\rm D}$, ${\rm ^4He}$, and ${\rm ^7Li}$, theoretical predictions
are known to be in reasonable agreements with the observations
\cite{BBN_review}.  As a result, we had an understanding of (some of)
the components in the universe.  This not only showed a very close
connection between high energy physics and cosmology, but also
provided a {\it quantitative} test of the big-bang scenario up to the
cosmic temperature of $\sim 1\ {\rm MeV}$.

Now, we are in a position to have an understanding of physics at the
electroweak scale.  Although most of the results from the high energy
experiments are more or less consistent with the predictions of the
standard model of the particle physics, many particle physicists are
expecting some new physics at the electroweak scale because of several
problems in the standard model, like the naturalness problem, the
hierarchy problem, and so on.  Among various possibilities, low-energy
supersymmetry (SUSY) is the prominent candidate of physics beyond the
standard model and hence signals from the superparticles are important
target of the future high energy experiments.

If the low-energy SUSY is realized in nature, it will also play
important roles in cosmology.  In particular, the lightest
superparticle (LSP) is known to be a good candidate of the cold dark
matter which, although there is no viable candidate within the
standard-model particles, is strongly suggested by the present precise
observations of the universe.  Indeed, recent results from the WMAP
suggests the dark matter density parameter to be
\cite{Bennett:2003bz,Eidelman:2004wy}
\begin{eqnarray}
  \Omega_{\rm DM}^{\rm (WMAP)} h^2 = 0.113^{+0.008}_{-0.009},
  \label{Omega_WMAP}
\end{eqnarray}
where $h$ here is the Hubble constant in units of $100\ {\rm
km/sec/Mpc}$.  In the framework of the low-energy supersymmetry,
however, it is not automatic to make the LSP to be the dark matter.
In addition, although thermal relic of the LSP is the usual candidate
of the dark matter, it has been also pointed out that the LSP dark
matter may have non-thermal origin \cite{non-thermal}.  Thus, once the
superparticles are found, it will be important to check if the thermal
relic of the LSP is really the dominant component of the cold dark
matter; if the theoretically calculated value of the LSP density well
agrees with the observed dark-matter density, 
then it will be an
important quantitative confirmation of the (simple) big-bang scenario
up to the freeze-out temperature of the LSP, $O(10\ {\rm GeV})$.

In order for a precise calculation of the relic density of the LSP, it
is necessary to determine the properties of the superparticles by
collider experiments.  In particular, the relic density is sensitive
to the properties of the LSP (such as the mass and couplings), precise
determination of those parameters will be necessary.  It has been
discussed that the high energy $e^+e^-$ linear collider, recently
named as the International Linear Collider (ILC), will be a good
facility for a detailed study of the superparticles; for various
cases, it has been shown that the $e^+e^-$ linear collider can
determine various mass and coupling parameters in the SUSY models very
well (see, for example, Refs.\ 
\cite{JLC1,NLC,Aguilar-Saavedra:2001rg}).  Although, in the current
situation, the LHC will probably be the first place where the
superparticles will be found, it may not be easy to determine the
SUSY parameters accurate enough to precisely calculate the relic
density of the LSP \cite{Allanach:2004xn}.  Thus, in the following, we
consider how the ILC can help to determine the relic density of the
LSP.

Relic density of the LSP depends on how the LSPs annihilate in the
early universe when they freeze out from the thermal bath.  The
dominant annihilation processes of the LSP are model-dependent and
there are several parameter regions where the relic density of the LSP
becomes consistent with the WMAP value, such as ``bulk region,''
``coannihilation region,'' ``rapid-annihilation funnels,''
``focus-point region,'' and so on \cite{LSPDM_WMAP}.  Thus, it is
necessary to study the individual cases.

In this letter, we consider one of the cases, the focus-point case; we
assume that the dark matter consists of the LSP from the focus-point
supersymmetric model, and address a question how and how well we can 
determine the density parameter of the LSP $\Omega_{\rm LSP}$ using 
the data from the ILC.  With reasonable assumptions, we will see that, 
in the focus point case, relic density of the LSP can be well constrained.
Before closing the introduction, we comment on how we proceed our
analysis.  We assume that the underlying model behind the minimal
supersymmetric standard model (MSSM) is the focus point model (with
grand unification) and that the dark matter is the thermal relic of the
LSP.  However, we will propose a procedure to calculate $\Omega_{\rm
LSP}$ without relying on any high-energy models (such as grand
unification, mSUGRA-type parameterization, and so on);  our aim
is to determine all the relevant {\it MSSM} parameters from the
collider experiments for the calculation of $\Omega_{\rm LSP}$.

We start with a brief review of the focus point scenario.  The
focus point scenario is characterized by the large (universal) scalar
mass at the grand unification theory (GUT) scale
\cite{focus}.\footnote
{Such a large universal scalar mass may be a result of the ``large
cut-off supergravity'' \cite{Ibe:2004mp}.}
Although all the squarks and sleptons (as well as the heavier Higgs
bosons) acquire multi-TeV masses, naturalness of the electroweak
symmetry breaking may be maintained as far as the gauginos and
Higgsinos remain relatively light.  Thus, in this scenario, gauginos
and Higgsinos are the ``light'' superparticles which may be accessible
with the ILC with center-of-mass energy lower than $\sim 1\ {\rm
TeV}$.

Since the important superparticles are the charginos and neutralinos,
the relevant parameters for our study are the $U(1)_Y$ and $SU(2)_L$
gaugino masses $m_{\rm G1}$ and $m_{\rm G2}$, the supersymmetric Higgs
mass $\mu_H$, and the ratio of the vacuum expectation values of two
Higgs bosons $\tan\beta$.  Properties of the charginos and the
neutralinos are determined by these parameters.  In particular, the
mass matrices of the charginos and the neutralinos are given
by\footnote
{In fact, the chargino and neutralino masses may acquire radiative
corrections, which should be taken into account in the actual study.
In the focus point scenario, dominant radiative corrections are from
the gauge boson loops and hence are calculable.  In our study, we
neglect the radiative corrections to the chargino and the neutralino
masses.}
\begin{eqnarray}
    {\cal M}_{\pm} &=& \left( \begin{array}{cc}
            - m_{\rm G2} & \sqrt{2} g_2 v \cos\beta \\
            - \sqrt{2} g_2 v \sin\beta & \mu_H
        \end{array} \right),
    \\
    {\cal M}_{0} &=& \left( \begin{array}{cccc}
            -m_{\rm G1} & 0 & - g_1 v \cos\beta & g_1 v \sin\beta \\
            0 & -m_{\rm G2} & g_2 v \cos\beta & - g_2 v \sin\beta \\
            - g_1 v \cos\beta & g_2 v \cos\beta & 0 & \mu_H \\
            g_1 v \sin\beta & - g_2 v \sin\beta & \mu_H & 0
        \end{array} \right),~~~
\end{eqnarray}
respectively.  Here, $g_1$ and $g_2$ are the gauge coupling constants
for $U(1)_Y$ and $SU(2)_L$, respectively, and $v\simeq 174\ {\rm GeV}$
is the (total) Higgs vacuum expectation value.  These mass matrices
are diagonalized by the unitary matrices $U_{\chi^+}$, $U_{\chi^-}$,
and $U_{\chi^0}$ as
\begin{eqnarray}
    U_{\chi^+}^T {\cal M}_{\pm} U_{\chi^-}
    = {\rm diag} (m_{\chi^\pm_1}, m_{\chi^\pm_2}),~~~
    U_{\chi^0}^T {\cal M}_{0} U_{\chi^0}
    = {\rm diag}
    (m_{\chi^0_1}, m_{\chi^0_2}, m_{\chi^0_3}, m_{\chi^0_4}).
\end{eqnarray}

If the GUT relation holds among the gaugino masses, the relation
$m_{\rm G2}\simeq 2m_{\rm G1}$ holds (at the electroweak scale) while,
for a successful electroweak symmetry breaking, $|\mu_H|$ becomes
larger than $|m_{\rm G1}|$.  Then, the lightest neutralino, which
becomes the LSP, is dominantly Bino although there is sizable
contamination of the Higgsino, as we mention below.

It should be noted that $\Omega_{\rm LSP}$ is also determined by the
previously mentioned four parameters: $m_{\rm G1}$, $m_{\rm G2}$,
$\mu_H$, and $\tan\beta$.  In the focus point scenario, the LSPs
dominantly annihilate into the $t\bar{t}$ pair and into the gauge
boson pairs ($\chi^0_1\chi^0_1\rightarrow W^+W^-$, $Z^0Z^0$) when the
LSPs freeze out from the thermal bath
\cite{Feng:2000gh,Allanach:2004xn,Ibe:2005jf}.  These processes are
through the Higgsino components in $\chi^0_1$; in the focus point
scenario, $|\mu_H|$ can be relatively close to $m_{\rm G1}$ and,
consequently, sizable contamination of the Higgsino is possible in
$\chi^0_1$.  In this case, $\Omega_{\rm LSP}$ may become consistent
with the WMAP value.  In the cosmologically interesting
parameter region where $\Omega_{\rm LSP}\sim 0.1$, $|\mu_H|$ becomes
smaller than $|m_{\rm G2}|$ and the lighter chargino $\chi^\pm_1$
becomes Higgsino-like while the heavier one $\chi^\pm_2$ becomes
Wino-like.  In addition, in the neutralino sector, $\chi^0_1$ (i.e.,
the LSP), $\chi^0_2$, $\chi^0_3$, and $\chi^0_4$ are Bino-like,
Higgsino-like, Higgsino-like, and Wino-like, respectively.  As a
result, the masses of $\chi^\pm_1$, $\chi^0_2$, and $\chi^0_3$ are
quite degenerate with the approximate relations $m_{\chi^0_1}\sim
|m_{\rm G1}|$, $m_{\chi^\pm_1}\sim m_{\chi^0_2}\sim
m_{\chi^0_3}\sim|\mu_H|$, and $m_{\chi^\pm_2}\sim m_{\chi^0_4}\sim
|m_{\rm G2}|$.  These facts will become very important in the
following study.

As we mentioned, $\Omega_{\rm LSP}$ is determined once the four
parameters $m_{\rm G1}$, $m_{\rm G2}$, $\mu_H$, and $\tan\beta$ are
fixed.  With the GUT relation, a narrow strip is obtained on the
$m_{\rm G2}$ vs.\ $\mu_H$ plane, where $\Omega_{\rm LSP}$ satisfies
the WMAP constraint Eq.(\ref{Omega_WMAP}). Such a strip is insensitive to
$\tan\beta$ (and the approximate relation $\mu_H\simeq 0.6m_{\rm G2}$
holds on such a strip when $m_{\rm G2}\gtrsim 300 \ {\rm GeV}$)
\cite{Ibe:2005jf}.  Although our strategy to determine $\Omega_{\rm
LSP}$ works for most of the points on the strip, (and even for cases
without the GUT relation), it will be instructive to see the detail of
several cases.  Thus, we pick up two parameter points where WMAP value
of $\Omega_{\rm LSP}$ is realized: Point 1 with relatively small
$m_{\rm G2}$ and Point 2 with larger $m_{\rm G2}$.  These points are
listed in Table \ref{table:params}; for these points, $\Omega_{\rm
LSP}$ and the lightest Higgs mass $m_h$ are calculated with adopting
that all the sfermion masses are $3\ {\rm TeV}$.  In the following,
several physical quantities such as the cross sections as well as the
estimated errors in the reconstructed $\Omega_{\rm LSP}$ will be given
for these points.  (In our numerical analysis, we use the DarkSUSY 
package \cite{Gondolo:2004sc} to calculate $\Omega_{\rm LSP}$.)

\begin{table}[t]
    \begin{center}
        \begin{tabular}{lcc}
            \hline\hline
            {} & {Point 1} & {Point 2} \\
            \hline
            $m_{\rm G1}$      & 144 GeV   & 240 GeV   \\
            $m_{\rm G2}$      & 300 GeV   & 500 GeV   \\
            $\mu_H$           & 200 GeV   & 307 GeV   \\ 
            $\tan\beta$       & 10        & 10 \\ 
            $m_{\chi^0_1}$    & 127 GeV   & 226 GeV   \\ 
            $m_{\chi^0_2}$    & 190 GeV   & 304 GeV   \\ 
            $m_{\chi^0_3}$    & 208 GeV   & 311 GeV   \\ 
            $m_{\chi^0_4}$    & 335 GeV   & 522 GeV   \\ 
            $m_{\chi^\pm_1}$  & 176 GeV   & 291 GeV   \\ 
            $m_{\chi^\pm_2}$  & 335 GeV   & 521 GeV   \\ 
            $m_h$             & 116 GeV   & 116 GeV   \\
            $\Omega_{\rm LSP}h^2$ & 0.113 & 0.113 \\
            \hline\hline
        \end{tabular}
        \caption{Points to be used for our analysis.}
        \label{table:params}
    \end{center}
\end{table}

Now we discuss the role of the ILC.  As we mentioned, in the focus
point case, the chargino(s) and neutralino(s) are the ones which can
be produced and studied at the ILC.  Thus, we focus on the question
what can be measured and studied by the production of the charginos
and the neutralinos.\footnote
{We assume that the squarks and sleptons will be known to be very
heavy by the study of the LHC, and also by the negative searches of
these particles at the ILC.}
For this purpose, we first calculate the production cross sections of
the charginos and neutralinos.  Since the sleptons are extremely
heavy, chargino and neutralino production processes are mediated by
the $s$-channel gauge-boson exchange diagrams; neglecting the
selectron diagrams, the chargino and neutralino production cross
sections are given by
\begin{eqnarray}
    \sigma (e^+ e_L^- \rightarrow \chi_X \chi_Y) &=& 
    \frac{N_{XY}\beta_{\rm f}}{8\pi} 
    \Bigg[ \left( \left| C_{LL} \right|^2 
        + \left| C_{LR} \right|^2 \right)
    \left( E_X E_Y + 
        \frac{1}{3} \left| {\bf p}_{\rm f} \right|^2 \right)
    \nonumber \\ &&
    + \left( C_{LL} C_{LR}^* + C_{LL}^* C_{LR} \right) m_X m_Y
    \Bigg],
\end{eqnarray}
where $E_X$, $E_Y$, and ${\bf p}_{\rm f}$ are the energies of $\chi_X$
and $\chi_Y$, and their three-momentum, respectively,
$N_{XY}=\frac{1}{2}$ when $\chi_X$ and $\chi_Y$ are identical and
$N_{XY}=1$ otherwise, and
\begin{eqnarray}
    \beta_{\rm f}^2 = \frac{1}{s^2}
    \left[ s^2 - 2 \left( m_X^2 + m_Y^2 \right) s
        + \left( m_X^2 - m_Y^2 \right)^2 \right],
\end{eqnarray}
with $\sqrt{s}$ being the center-of-mass energy of the ILC.  Here, the
polarization of the electron is specified to be left-handed while the
average over the positron polarization is taken.  Result for the
right-handed electron is given by a similar formula by replacing
$C_{LL}\rightarrow C_{RL}$ and $C_{LR}\rightarrow C_{RR}$.  For the
chargino production,
\begin{eqnarray}
    \left[ C_{LL} \right]_{\chi^\pm_X \chi^\mp_Y} = 
    \frac{e^2}{s} \delta_{XY} 
    + \frac{g_Z^{e_L}}{s-m_Z^2} \left( -\frac{g_2^2}{g_Z} 
        \left[ U_{\chi^-} \right]_{1X}^*
        \left[ U_{\chi^-} \right]_{1Y}
        + \frac{g_1^2 - g_2^2}{2g_Z} 
        \left[ U_{\chi^-} \right]_{2X}^*
        \left[ U_{\chi^-} \right]_{2Y}
    \right),
\end{eqnarray}
where $g_Z^2\equiv g_1^2+g_2^2$, and $e=(g_1^{-2}+g_2^{-2})^{-1/2}$ is
the electric charge.  $\left[ C_{RL} \right]_{\chi^\pm_X \chi^\mp_Y}$
is obtained by replacing $g_Z^{e_L}\rightarrow g_Z^{e_R}$, where
\begin{eqnarray}
    g_Z^{e_L} \equiv \frac{g_1^2 - g_2^2}{2g_Z},~~~
    g_Z^{e_R} \equiv \frac{g_1^2}{g_Z}.
\end{eqnarray}
In addition, $\left[ C_{LR} \right]_{\chi^\pm_X \chi^\mp_Y}$ and
$\left[ C_{RR} \right]_{\chi^\pm_X \chi^\mp_Y}$ are obtained by
the following replacements:
\begin{eqnarray}
    \left[ C_{LR} \right]_{\chi^\pm_X \chi^\mp_Y} =
    \left. \left[ C_{LL} \right]_{\chi^\pm_X \chi^\mp_Y} 
    \right|_{U_{\chi^-}\rightarrow U_{\chi^+}},~~~
    \left[ C_{RR} \right]_{\chi^\pm_X \chi^\mp_Y} =
    \left. \left[ C_{RL} \right]_{\chi^\pm_X \chi^\mp_Y} 
    \right|_{U_{\chi^-}\rightarrow U_{\chi^+}}.
\end{eqnarray}
For the neutralino production processes,
\begin{eqnarray}
    \left[ C_{LL} \right]_{\chi^0_X \chi^0_Y} =
    \frac{g_Z^{e_L} g_Z}{2(s-m_Z^2)} 
    \left(
        \left[ U_{\chi^0} \right]_{3X}^*
        \left[ U_{\chi^0} \right]_{3Y}
        - \left[ U_{\chi^0} \right]_{4X}^*
        \left[ U_{\chi^0} \right]_{4Y}
    \right),
\end{eqnarray}
and $\left[ C_{LR} \right]_{\chi^0_X \chi^0_Y}= - \left[ C_{LL}
\right]_{\chi^0_Y \chi^0_X}$.  For the right-polarized electron, the
results are obtained by replacing $g_Z^{e_L}\rightarrow g_Z^{e_R}$.

\begin{figure}[t]
    \centerline{\epsfxsize=0.5\textwidth\epsfbox{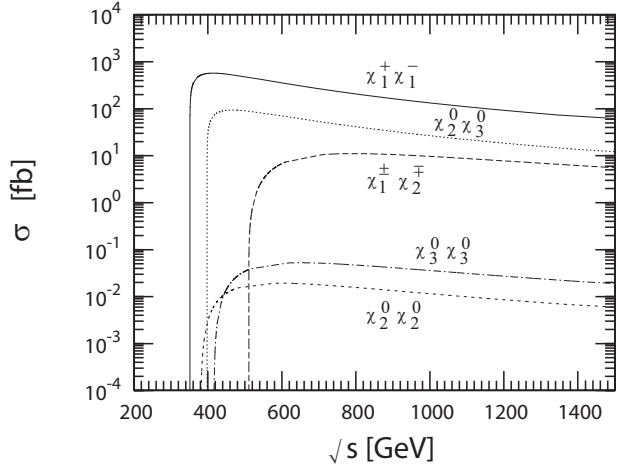}}
    \caption{Production cross sections of the charginos and neutralinos
     for Point 1 for the processes  as functions of $\sqrt{s}$.  Here, 
    we have neglected the selectron-exchange diagrams and averaged over 
    the polarization of the electron beam.}
    \label{fig:cs1}
\end{figure}

\begin{figure}[t]
    \centerline{\epsfxsize=0.5\textwidth\epsfbox{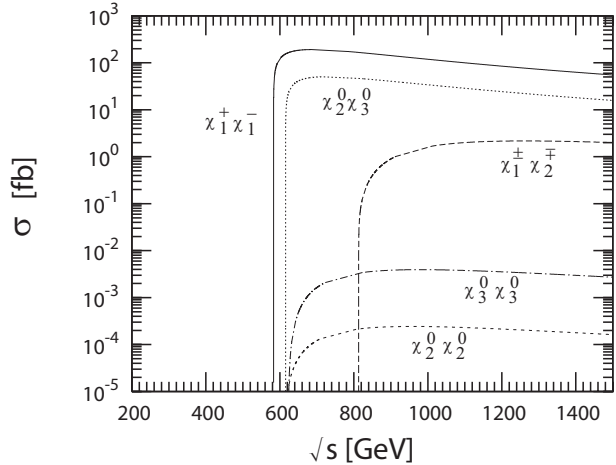}}
    \caption{Same as Fig.\ \ref{fig:cs1}, except for Point 2.}
    \label{fig:cs2}
\end{figure}

To see the size of the cross sections, we plot the cross sections for
various final states in Figs.\ \ref{fig:cs1} and \ref{fig:cs2} for
Point 1 and Point 2, respectively, as functions of $\sqrt{s}$.  As one
can see, for the processes $e^+e^-\rightarrow\chi^+_1\chi^-_1$ and
$\chi^0_2\chi^0_3$, cross sections become significantly large (if
these processes are kinematically allowed).  As we will see, these
processes give very important information in calculating the relic
density of the LSP.\footnote
{Notice that the chargino $\chi^\pm_1$ mostly decays into
$W^{\pm(*)}\chi^0_1$ final state (where the superscript $*$ is for
virtual particles) while $\chi^0_2$ and $\chi^0_3$ decay into
$Z^{0(*)}\chi^0_1$ and $h\chi^0_1$ final states (where $h$ is the
lightest Higgs boson).  Thus, we expect that some of the chargino and
neutralino production events can be distinguished by selecting
particular final states.  For example, for
$e^+e^-\rightarrow\chi^+_1\chi^-_1$, we can use $1~l+2~{\rm jets}+{\rm
missing}~E_{\rm T}$ events, which is not from $\chi^0_2\chi^0_3$
production, and for $e^+e^-\rightarrow\chi^0_2\chi^0_3$, $2~l+2~{\rm
jets}+{\rm missing}~E_{\rm T}$ events can be used.}
On the contrary, cross sections $\sigma
(e^+e^-\rightarrow\chi^0_2\chi^0_2)$ and $\sigma
(e^+e^-\rightarrow\chi^0_3\chi^0_3)$ are very small.  This is due to
the facts that the cross sections acquire suppressions from the mixing
factors and that these are $p$-wave processes.  Note also that the
measurements of the cross sections for
$e^+e^-\rightarrow\chi^+_1\chi^-_1$ and $\chi^0_2\chi^0_3$ will tell
us that $\chi^\pm_1$, $\chi^0_2$, and $\chi^0_3$ are Higgsino-like,
not Wino-like.

A great advantage of the ILC is that the beam energy can be easily
tuned.  Thus, it is possible to study various production processes at
the threshold region.  In particular, in the case of the focus-point
scenario, threshold scans of the chargino and neutralino productions
give us important information.  From the chargino production process
$e^+e^-\rightarrow\chi^+_1\chi^-_1$, we can have a precise measurement
of the (lighter) chargino mass $m_{\chi^\pm_1}$.  Study at the region
$\sqrt{s}\sim m_{\chi^0_2}+m_{\chi^0_3}$ is also interesting.  At
$\sqrt{s}\sim 2|\mu_H|$, three different neutralino production
processes become kinematically allowed since the masses of $\chi^0_2$
and $\chi^0_3$ are close.  However, the cross sections for
$e^+e^-\rightarrow\chi^0_2\chi^0_2$ and
$e^+e^-\rightarrow\chi^0_3\chi^0_3$ are very small in particular at
the threshold region since these processes are $p$-wave suppressed; we
can see that the cross sections for these processes are much smaller
than $\sigma (e^+e^-\rightarrow\chi^0_2\chi^0_3)$.  Thus, we observe
only the process $e^+e^-\rightarrow\chi^0_2\chi^0_3$.  Since the
process $e^+e^-\rightarrow\chi^0_2\chi^0_3$ becomes kinematically
allowed when $\sqrt{s}=m_{\chi^0_2}+m_{\chi^0_3}$, we can determine
the combination $m_{\chi^0_2}+m_{\chi^0_3}$ or, equivalently, the
averaged mass of $\chi^0_2$ and $\chi^0_3$:
\begin{eqnarray}
    \bar{m}_{\chi^0_{23}} \equiv \frac{1}{2}
    \left( m_{\chi^0_2} + m_{\chi^0_3}\right),
\end{eqnarray}
by the scan around $\sqrt{s}\sim m_{\chi^0_2}+m_{\chi^0_3}$.

In order to study the properties of the LSP, on the contrary,
$\sqrt{s}$ should better be optimized so that the production of the
lighter chargino, which decays into the LSP $\chi^0_1$, is enhanced.
Once the charginos are copiously produced, then the mass of the LSP
(more precisely, the mass difference $m_{\chi^\pm_1}-m_{\chi^0_1}$) is
determined by the study of the energy distribution of the decay
products.\footnote
{From the $\chi^0_2\chi^0_3$ production, we may in principle perform a
similar analysis to determine the mass of the LSP.  In this case,
however, error may become larger because the masses of $\chi^0_2$ and
$\chi^0_3$ differ by $O(1\ {\rm GeV})$.  Global fit using all the data
should be appropriate once the charginos and neutralinos are really
found in the future.}

At the ILC, errors in the measurements of $m_{\chi^\pm_1}$,
$\bar{m}_{\chi^0_{23}}$, and $m_{\chi^0_1}$ are expected to be mostly
from the detector resolutions \cite{Aguilar-Saavedra:2001rg}.  For
example, it was pointed out that, for some choice of the SUSY
parameters, masses of the charginos can be determined using $e^+e^-$
colliders with the errors of $\sim 50\ {\rm MeV}$ by the threshold
scan.  In addition, from the energy distribution of the decay products
of the chargino and neutralinos, the mass of the LSP is also
determined with the uncertainty of $\sim 50\ {\rm MeV}$.  Although
these results are for the case of Wino-like chargino and neutralino,
we expect that three mass parameters (i.e., $m_{\chi^\pm_1}$,
$m_{\chi^0_1}$, and $\bar{m}_{\chi^0_{23}}$) are accurately measured
once $\chi^\pm_1$, $\chi^0_2$, and $\chi^0_3$ become kinematically
accessible at the ILC.  Since $\chi^0_1$ is Bino-like while
$\chi^\pm_1$ (as well as $\chi^0_2$ and $\chi^0_3$) are Higgsino-like,
we can constrain $m_{\rm G1}$ and $\mu_H$ from the measurements of
$m_{\chi^0_1}$ and $m_{\chi^\pm_1}$ (or from the masses of other
Higgsino-like neutralinos).  In addition, the ``mass difference''
$\bar{m}_{\chi^0_{23}}-m_{\chi^\pm_1}$ is sensitive to some
combination of $\tan\beta$ and $m_{\rm G2}$.

Although the relic density of the LSP depends on four MSSM parameters
($m_{\rm G1}$, $m_{\rm G2}$, $\mu_H$, and $\tan\beta$), interesting
bound on $\Omega_{\rm LSP}$ can be obtained even at this stage.  To
see this, we can perform the following analysis.  Let us imagine a
situation where $m_{\chi^\pm_1}$, $m_{\chi^0_1}$, and
$\bar{m}_{\chi^0_{23}}$ are well measured at the ILC.  Using these
quantities, we impose three constraints on the four underlying
parameters and determine $m_{\rm G1}$, $\mu_H$, and $\tan\beta$ as
functions of $m_{\rm G2}$.  In the determination of $m_{\rm G1}$
and $\mu_H$, in fact, there are four possible choices of their signs:
$({\rm sign}(m_{\rm G1}), {\rm sign}(\mu_H))=(+,+)$, $(+,-)$, $(-,+)$,
and $(-,-)$.\footnote
{To be more precise, these signs are the relative signs between
$m_{\rm G1}$ and $m_{\rm G2}$ or $\mu_H$ and $m_{\rm G2}$.  We assume
that the gaugino masses and $\mu_H$ are real in order to avoid
constraints from CP violations.}
Effects of the signs of $\mu_H$ and $m_{\rm G1}$ are quite different.
In order to see how the reconstructed relic density depends on $m_{\rm
G2}$ and ${\rm sign}(\mu_H)$, here let us consider the case where the
sign of the reconstructed $m_{\rm G1}$ is the same as that of the
underlying one; effects of ${\rm sign}(m_{\rm G1})$ will be discussed
later.  Once we reconstruct $m_{\rm G1}$, $\mu_H$, and $\tan\beta$, we
calculate the relic density of the LSP as a function of $m_{\rm G2}$,
which we call
\begin{eqnarray*}
    \hat{\Omega}_{\rm LSP} 
    (m_{\rm G2}; m_{\chi^\pm_1}, m_{\chi^0_1}, 
    \bar{m}_{\chi^0_{23}}).
\end{eqnarray*}

\begin{figure}[t]
    \centerline{\epsfxsize=0.5\textwidth\epsfbox{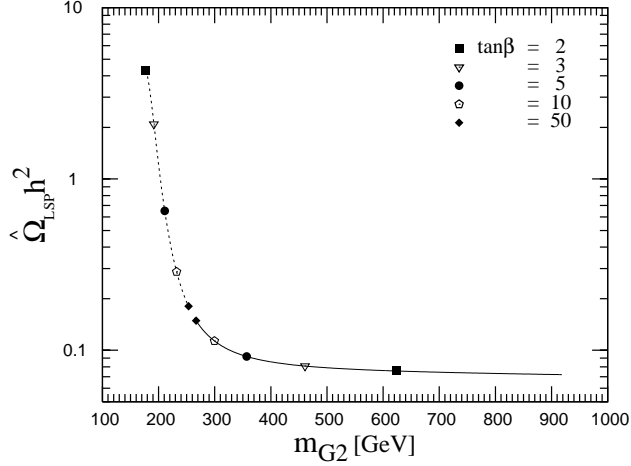}}
    \caption{
    $\hat{\Omega}_{\rm LSP} (m_{\rm G2}; m_{\chi^\pm_1},
    m_{\chi^0_1}, \bar{m}_{\chi^0_{23}})$ as a function of $m_{\rm
    G2}$, where $m_{\chi^\pm_1}$, $m_{\chi^0_1}$, and
    $\bar{m}_{\chi^0_{23}}$ are fixed by the underlying values for
    Point 1 with positive $\mu_H$ (solid) and negative $\mu_H$
 (dashed). Marks on the figure indicate the points  with
   $\tan\beta=2,3,5,10,50$.
    }
    \label{fig:omega1}
\end{figure}

\begin{figure}[t]
    \centerline{\epsfxsize=0.5\textwidth\epsfbox{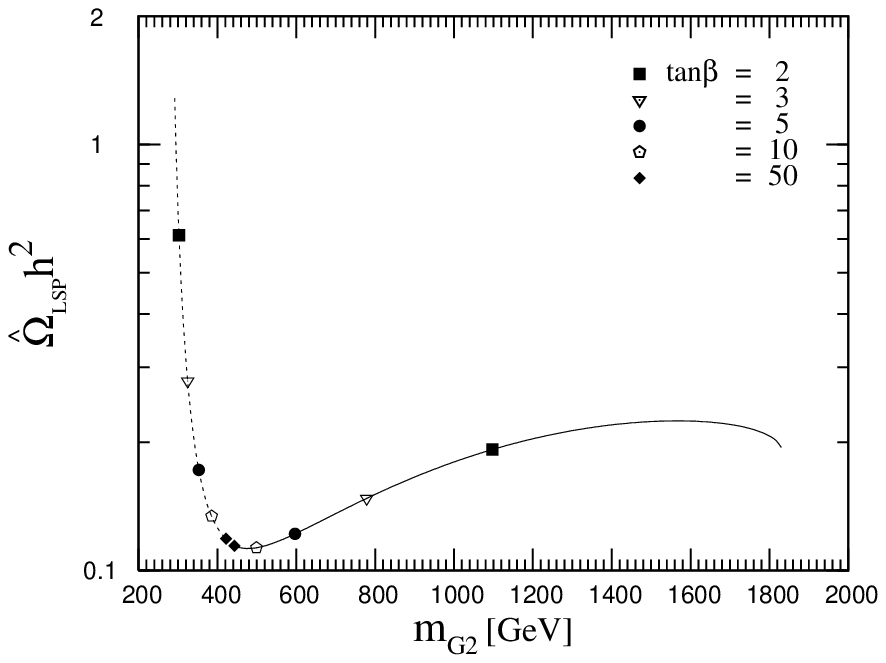}}
    \caption{ Same as Fig.\ \ref{fig:omega1}, except for Point 2.
    }
    \label{fig:omega2}
\end{figure}

In Figs.\ \ref{fig:omega1} and \ref{fig:omega2}, we plot
$\hat{\Omega}_{\rm LSP} (m_{\rm G2}; m_{\chi^\pm_1}, m_{\chi^0_1},
\bar{m}_{\chi^0_{23}})$ as a function of $m_{\rm G2}$, with
$m_{\chi^\pm_1}$, $m_{\chi^0_1}$, and $\bar{m}_{\chi^0_{23}}$ being
fixed to be the values from Point 1 and Point 2, respectively. We use
a fixed value of the lightest Higgs boson mass $m_h$. (We have checked
that $\hat{\Omega}_{\rm LSP}$ is insensitive to the change of $m_h$.)
The lines have endpoints; this is due to the fact that, when $m_{\rm
G2}$ becomes too large or too small, there is no value of $\tan\beta$
which consistently reproduces the observed mass spectrum.  To
demonstrate this, we also showed the points where $\tan\beta$ takes
several specific values.  
Here, one should note that there exists two-fold ambiguity from
the sign of $\mu_H$; in our study, we reconstructed $\mu_H$ for two
cases (with the sign of the underlying value of $\mu_H$ being fixed). 
In the figures, the results
for the cases with positive and negative $\mu_H$ are shown.  
As one
can see, the line for the $\mu_H<0$ case is ``attached'' to one of the
endpoints of the line for $\mu_H>0$.  This is from the fact that, at
the tree level, flip of ${\rm sign}(\mu_H)$ is equivalent to the
change $\beta\rightarrow\pi -\beta$.

As one can see, the case with negative $\mu_H$ may give large
uncertainty in the reconstructed $\Omega_{\rm LSP}$.  If $\mu_H<0$,
however, smaller $m_{\rm G2}$ is required than in the case of
positive-$\mu_H$ in order to reproduce the observed mass spectrum, as
shown in the figures.  Then, $m_{\chi^\pm_2}$, for example, may become
smaller than the experimental bound from the negative search for the
$\chi^\pm_1\chi^\mp_2$ production process.  For Point 1 (Point 2),
$\sqrt{s}\gtrsim 480\ {\rm GeV}$ ($750\ {\rm GeV}$) is enough to
exclude the $\mu_H<0$ case.  Thus, in the following, we assume that
this is the case and neglect the uncertainty from the $\mu_H<0$ case.
Then, even without any further constraint on $m_{\rm G2}$, relic
density of the LSP can be determined within a factor of $\sim 2$ or
smaller.

For a quantitative study of the uncertainties in the reconstructed LSP
density, we define the following quantity:
\begin{eqnarray}
    D_{m_I} \equiv
    \left|\frac{\partial \ln \hat{\Omega}_{\rm LSP}}
    {\partial \ln m_I}\right|, ~~~ 
    m_I = m_{\chi^\pm_1}, m_{\chi^0_1}, \bar{m}_{\chi^0_{23}},
    m_{\rm G2},
\end{eqnarray}
so that\footnote
{In fact, some of the observables may be correlated (in particular
because, as we mentioned, one of the observable would be
$m_{\chi^\pm_1}-m_{\chi^0_1}$, not $m_{\chi^0_1}$).  Here, we neglect
effects of such correlations.}
\begin{eqnarray}
    \frac{\delta \hat{\Omega}_{\rm LSP}}{\hat{\Omega}_{\rm LSP}}
    =
   \left[ 
    \left(D_{m_{\chi^\pm_1}} \frac{\delta
     m_{\chi^\pm_1}}{m_{\chi^\pm_1}}
    \right)^2
    + \left(D_{m_{\chi^0_1}} \frac{\delta m_{\chi^0_1}}{m_{\chi^0_1}}
    \right)^2
    +\left(D_{\bar{m}_{\chi^0_{23}}} 
    \frac{\delta \bar{m}_{\chi^0_{23}}}{\bar{m}_{\chi^0_{23}}}
    \right)^2
    + \left(D_{m_{\rm G2}}
    \frac{\delta m_{\rm G2}}{m_{\rm G2}}
    \right)^2
\right]^{1/2}.
\end{eqnarray}
For Point 1 and Point 2, we found $(D_{m_{\chi^\pm_1}},
D_{m_{\chi^0_1}}, D_{\bar{m}_{\chi^0_{23}}}, D_{m_{\rm G2}})$ to be
$(19, 8.1, 8.6, 1.5)$ and $(7.5, 7.9, 8.5, \sim 0.01)$, respectively.  
(Here, notice that $D_{m_{\rm G2}}$ for Point 2 is accidentally small
because of the choice of the underlying parameters.) 
As we mentioned, at the ILC, $m_{\chi^\pm_1}$, $m_{\chi^0_1}$, and
$\bar{m}_{\chi^0_{23}}$ will be well determined with the error of
$0.1\ \%$ or less \cite{Aguilar-Saavedra:2001rg} (although the errors
should depend on the model parameters).  Using the values of
$D_{m_{\chi^\pm_1}}$, $D_{m_{\chi^0_1}}$, and
$D_{\bar{m}_{\chi^0_{23}}}$ given above, $\delta\hat{\Omega}_{\rm
LSP}$ from $\delta m_{\chi^\pm_1}$, $\delta m_{\chi^0_1}$, and
$\delta\bar{m}_{\chi^0_{23}}$ is very small.

For a better determination of $\Omega_{\rm LSP}$, some
information about $m_{\rm G2}$ is necessary.  For example, if we try
to determine $\Omega_{\rm LSP}$ with $\sim 10\ \%$ accuracy, which is
the level of the WMAP value, $m_{\rm G2}$ should be determined within
the uncertainty of $\sim 30\ {\rm GeV}$ for Point 1 and 
$\sim 170\ {\rm GeV}$
for Point 2.  If we impose the GUT relation among the gaugino masses,
$m_{\rm G2}$ is accurately determined.  However, we do not
pursue this direction since we hope to calculate  $\Omega_{\rm
LSP}$ in the framework of the MSSM.

The $SU(2)_L$ gaugino mass $m_{\rm G2}$ is approximately equal to the
masses of $\chi^\pm_2$ and $\chi^0_4$.  Thus, if we know something
about $m_{\chi^\pm_2}$ or about $m_{\chi^0_4}$, it will help us to set
a bound on $m_{\rm G2}$.  Some constraint may be given by the LHC.
Even in the focus-point scenario, a sizable amount of the gluino may
be produced at the LHC, and some of them decay into $\chi^\pm_2$ or
$\chi^0_4$ (and other quark jets).  These charginos and neutralinos
will decay by emitting $W^\pm$ or $Z^0$ boson (or the lightest Higgs
boson, in some case).  If the gauge bosons subsequently decay into the
leptons, for example, we may have events with multi-leptons and
missing $E_{\rm T}$, whose background is probably relatively small.
By studying the energy distribution of the charged leptons, some
information about the Wino-like particles may be obtained.

If $\chi^\pm_2$ or $\chi^0_4$ becomes kinematically accessible at the
ILC, some direct information of these particles will become available.
For example, we can look for the associated production processes of
these Wino-like particles with lighter charginos or neutralinos
(although the cross sections for the associated productions are
relatively suppressed).  For example, with large enough $\sqrt{s}$,
cross section for the $\chi^\pm_1\chi^\mp_2$ production process is
sizable, as shown in Figs.\ \ref{fig:cs1} and \ref{fig:cs2}. With the
threshold scan of this process, $m_{\chi^\pm_2}$ may be measured,
resulting in a good determination of $m_{\rm G2}$. 

In order to constrain $m_{\rm G2}$, we may also use $\tan\beta$
dependence of the lightest Higgs mass.  As shown in Figs.\ 
\ref{fig:omega1} and \ref{fig:omega2}, the reconstructed value of 
$\tan\beta$ depends on $m_{\rm G2}$.  Importantly, smaller value of 
$\tan\beta$ results in more suppressed value of the lightest Higgs mass.  
Although the lightest Higgs mass depends on other MSSM parameters (in
particular, on the stop masses \cite{SUSYHiggs}), too small
$\tan\beta$ will be excluded by the detailed study of the lightest
Higgs, which will be possible at the ILC
\cite{JLC1,NLC,Aguilar-Saavedra:2001rg}.  If we obtain a lower bound
on $\tan\beta$ from the measurement of the Higgs mass, it will also
help to determine the relic density of the LSP.

Another possibility to constrain $m_{\rm G2}$ is to measure the cross
sections for the processes $e^+e^-\rightarrow\chi^0_1\chi^0_2$ and
$e^+e^-\rightarrow\chi^0_1\chi^0_3$.  In some case, $\chi^0_2$ and
$\chi^0_3$ both dominantly decay into $Z^{0(*)}\chi^0_1$.  If so, the
$\chi^0_1\chi^0_2$ and $\chi^0_1\chi^0_3$ production processes has the
final state with $2~l+{\rm missing}~E_{\rm T}$ and $2~{\rm jets}+{\rm
missing}~E_{\rm T}$.  It may not be easy to distinguish these two
events, but we can just consider the total cross section
$\sigma(e^+e^-\rightarrow\chi^0_1\chi^0_2)+
\sigma(e^+e^-\rightarrow\chi^0_1\chi^0_3)$.  For Point 2, for example,
the total cross section $\sigma(e^+e^-\rightarrow\chi^0_1\chi^0_2)+
\sigma(e^+e^-\rightarrow\chi^0_1\chi^0_3)$ monotonically 
increases as a function of $m_{\rm G2}$ from $15\ {\rm
fb}$ (for $m_{\rm G2}=440\ {\rm GeV}$) to $29\ {\rm fb}$ (for $m_{\rm
G2}=1830\ {\rm GeV}$).  Thus, if this cross section is precisely
measured, we can obtain another information about $m_{\rm G2}$.
However, it should be noted that several serious standard-model
backgrounds may exist.  In particular, the processes
$e^+e^-\rightarrow W^+W^-$, $\nu_e\bar{\nu}_eZ^0$, and $W^+W^-Z^0$
have large cross sections.  (For $\sqrt{s}=1\ {\rm TeV}$,
$\sigma(e^+e^-\rightarrow W^+W^-)\sim 3\ {\rm pb}$,
$\sigma(e^+e^-\rightarrow\nu_e\bar{\nu}_eZ^0)\sim 900\ {\rm fb}$, and
$\sigma(e^+e^-\rightarrow W^+W^-Z^0)\sim 60\ {\rm fb}$ for the 
unpolarized electron beam \cite{Murayama:1991uq}.)  Although these 
processes are suppressed for the right-polarized electron beam, 
cross sections for these processes are much larger than
$\sigma(e^+e^-\rightarrow\chi^0_1\chi^0_2)+
\sigma(e^+e^-\rightarrow\chi^0_1\chi^0_3)$.  Thus, some appropriate
kinematical cuts should be necessary to use these modes for the
determination of $\sigma(e^+e^-\rightarrow\chi^0_1\chi^0_2)+
\sigma(e^+e^-\rightarrow\chi^0_1\chi^0_3)$.

So far, we have not considered effects of the sign of $m_{\rm G1}$.
Unfortunately, $\Omega_{\rm LSP}$ depends on the sign of $m_{\rm G1}$
although the determination of ${\rm sign}(m_{\rm G1})$ seems
challenging.  For the case with negative $m_{\rm G1}$, $\Omega_{\rm
LSP}h^2$ varies from 0.9 to 7.4 (Point 1) and from 0.2 to 6.7 (Point
2).  If ${\rm sign}(m_{\rm G1})$ is undetermined, thus, two-fold
ambiguity will remain. However, experimental determination of ${\rm
sign}(m_{\rm G1})$ may be possible \cite{Choi:2001ww}.  In addition,
if the GUT relation among the (absolute values of) gaugino masses is
experimentally confirmed, it will give another hint of the signs of
the gaugino masses.

We have not discussed possible errors of $\Omega_{\rm LSP}$
originating from parameters other than $m_{\chi^\pm_1}$,
$m_{\chi^0_1}$, and $\bar{m}_{\chi^0_{23}}$, and $m_{\rm G2}$.
Theoretical prediction on $\Omega_{\rm LSP}$, in fact, depends also on
other parameters.  For the process $\chi^0_1\chi^0_1\rightarrow
t\bar{t}$, which can be the dominant pair annihilation process of the
LSP, there exists the $t$ channel stop exchange diagram, so the cross
section for this process depends on the stop masses.  Since the stops
are very heavy in the focus point scenario, the stop masses will not
be directly measured at the ILC.  In addition, the cross section for
this process also depends on the lightest Higgs boson mass through the
$s$-channel exchange of the Higgs boson.  (Of course, the lightest
Higgs boson mass will be accurately measured at the ILC, so the latter
diagram will not give serious uncertainty.)  However, the process
$\chi^0_1\chi^0_1\rightarrow t\bar{t}$ is dominated by the $s$-channel
$Z^0$ exchange diagram \cite{Allanach:2004xn}.  We have checked that
the reconstructed value of $\Omega_{\rm LSP}$ is insensitive to the
choice of the stop mass and $m_h$.  Other possible uncertainty may be
from the pair annihilation into the $b\bar{b}$ final state, which may
give $\sim 10\ \%$ contribution to the total annihilation cross
section when $\tan\beta$ is very as large as $\sim 50$ because of the
enhanced bottom-quark Yukawa coupling \cite{Allanach:2004xn}.  The
cross section for the process $\chi^0_1\chi^0_1\rightarrow b\bar{b}$
cannot be calculated unless we know the pseudo-scalar Higgs boson
mass.  Thus, $\sim 10\%$ uncertainity may remain unless the
pseudo-scalar Higgs boson mass is somehow constrained.  Uncertainties
from the first and second generation sfermions are irrelevant since
the pair annihilation into light fermions is $p$-wave suppressed.

In this letter, we have seen that, if the dominant component of the
cold dark matter is the thermal relics of the LSP in the focus-point
supersymmetric model, we will have a good chance to theoretically
reconstruct the cold dark matter density once the superparticles
becomes kinematically accessible at the ILC.  
Positive confirmation of the dark matter density will give us a
quantitative understanding of our universe up to the freeze-out
temperature of the LSP, $O(10\ {\rm GeV})$.
Of course,
even in the framework of the supersymmetric models, there are several
possibilities to realize the LSP dark matter.  Superparticle spectrum
and the observables at the collider experiments depend on the model.
Thus, for other cases, it is also necessary to study how and how well
we can reconstruct the dark matter density.  In addition, in our
study, we have not performed detailed detector simulations to have
accurate errors in the observables (although we have seen that, using
the reasonable estimate of the errors, determination of $\Omega_{\rm
LSP}$ seems promising with a good accuracy).  We leave these for
future studies.

{\it Note added:} While finalizing this letter, we found the paper
\cite{BaeMusPar} which may have some relevance with our analysis.

{\it Acknowledgment:} This work  is supported in part by the 21st 
century COE program,
``Exploring New Science by Bridging Particle-Matter Hierarchy.''
The work of T.M. is also supported by the Grants-in Aid of the Ministry of
Education, Science, Sports, and Culture of Japan No.\ 15540247.


\begin{thebibliography}{99}

\bibitem{BBN_Gamow}
    G.~Gamow,
    Phys.\ Rev.\ {\bf 70} (1946) 572;
    R.~A.~Alpher, H.~Bethe, and G.~Gamow,
    Phys.\ Rev.\ {\bf 73} (1948) 803.

\bibitem{BBN_review}
    For review, see, for example,
    B.~D.~Fields and S.~Sarkar,
    Phys.\ Lett.\ B {\bf 592} (2004) 202.

\bibitem{Bennett:2003bz}
    C.~L.~Bennett {\it et al.},
    Astrophys.\ J.\ Suppl.\  {\bf 148} (2003) 1.

\bibitem{Eidelman:2004wy}
    S.~Eidelman {\it et al.}  [Particle Data Group],
    Phys.\ Lett.\ B {\bf 592} (2004) 1.

\bibitem{non-thermal}
    T.~Moroi and L.~Randall,
    Nucl.\ Phys.\ B {\bf 570} (2000) 455;
    M.~Fujii and K.~Hamaguchi,
    Phys.\ Rev.\ D {\bf 66} (2002) 083501.

\bibitem{JLC1}
    S.~Matsumoto {\it et al.} [JLC Group],
    ``JLC-1,'' 
    KEK Report 92-16 (1992).

\bibitem{NLC}
    S.~Kuhlman {\it et al.} [The NLC ZDR Design Group and The NLC
    Physics Working Group],
    ``Physics and Technology of the Next Linear Collider,''
    BNL 52-502 (1996).

\bibitem{Aguilar-Saavedra:2001rg}
    J.~A.~Aguilar-Saavedra {\it et al.}  
    [ECFA/DESY LC Physics Working Group],
    arXiv:hep-ph/0106315.

\bibitem{Allanach:2004xn}
    B.~C.~Allanach, G.~Belanger, F.~Boudjema and A.~Pukhov,
    JHEP {\bf 0412} (2004) 020.

\bibitem{LSPDM_WMAP}
    J.~R.~Ellis, K.~A.~Olive, Y.~Santoso and V.~C.~Spanos,
    Phys.\ Lett.\ B {\bf 565} (2003) 176;
    H.~Baer and C.~Balazs,
    JCAP {\bf 0305} (2003) 006;
    U.~Chattopadhyay, A.~Corsetti and P.~Nath,
    Phys.\ Rev.\ D {\bf 68} (2003) 035005.

\bibitem{focus}
    J.~L.~Feng and T.~Moroi,
    Phys.\ Rev.\ D {\bf 61} (2000) 095004;
    J.~L.~Feng, K.~T.~Matchev and T.~Moroi,
    Phys.\ Rev.\ Lett.\  {\bf 84} (2000) 2322;
    Phys.\ Rev.\ D {\bf 61} (2000) 075005.

\bibitem{Ibe:2004mp}
    M.~Ibe, K.~I.~Izawa and T.~Yanagida,
    Phys.\ Rev.\ D {\bf 71} (2005) 035005.
    
\bibitem{Feng:2000gh}
    J.~L.~Feng, K.~T.~Matchev and F.~Wilczek,
    Phys.\ Lett.\ B {\bf 482} (2000) 388.

\bibitem{Ibe:2005jf}
    M.~Ibe, T.~Moroi and T.~Yanagida,
    arXiv:hep-ph/0502074.

\bibitem{Gondolo:2004sc}
    P.~Gondolo {\it et al.},
    JCAP {\bf 0407} (2004) 008.

\bibitem{SUSYHiggs}
    Y.~Okada, M.~Yamaguchi and T.~Yanagida,
    Prog.\ Theor.\ Phys.\  {\bf 85} (1991) 1;
    H.~E.~Haber and R.~Hempfling,
    Phys.\ Rev.\ Lett.\  {\bf 66} (1991) 1815;
    J.~R.~Ellis, G.~Ridolfi and F.~Zwirner,
    Phys.\ Lett.\ B {\bf 257} (1991) 83.

\bibitem{Murayama:1991uq}
    H.~Murayama,
    Ph.\ D.\ Thesis (UT-580).

\bibitem{Choi:2001ww}
    S.~Y.~Choi, J.~Kalinowski, G.~Moortgat-Pick and P.~M.~Zerwas,
    Eur.\ Phys.\ J.\ C {\bf 22} (2001) 563
    [Addendum-ibid.\ C {\bf 23} (2002) 769].

\bibitem{BaeMusPar}
    H.~Baer, A.~Mustafayev, E.-K.~Park and S.~Profumo,
    arXiv:hep-ph/0505227.

\end{thebibliography}
\end{document}